\documentclass{elsart}

\usepackage{amsmath,amsfonts,amssymb}
\usepackage[pdftex]{graphicx}
\usepackage{multimedia}
\usepackage{graphicx}
\DeclareGraphicsExtensions{.png,.pdf,.mps}

\usepackage{amssymb}
\usepackage[english]{babel}
\usepackage{graphicx}
\usepackage{textcomp}
\usepackage{amsmath}

\numberwithin{equation}{section}
\def\sign{\mathop{\rm sgn}\nolimits}
\usepackage{amssymb}
\newtheorem{theorem}{Theorem}[section]
\newtheorem{lemma}{Lemma}[section]
\newtheorem{remark}{Remark}[section]

\begin{document}

\begin{frontmatter}

\title{Mathematical model of influence of  friction on the vortex motion}


\author[1]{Olga S. Rozanova\corauthref{cor1} }
\ead{rozanova@mech.math.msu.su} \corauth[cor1]{Corresponding author}
\author[2]{Jui-Ling Yu}
\author[3]{Chin-Kun Hu}

\address[1]{Department of Mechanics and Mathematrics, Moscow
State University, Moscow 119992 Russia}
\address[2]{Department of Financial and Computational Mathematics,
Providence University, Taichung, 43301, Taiwan\\ and National Center
of Theoretical Sciences at Taipei, Physics Division, National Taiwan
University, Taipei 10617, Taiwan}
\address[3]{Institute of Physics, Academia Sinica, Nankang, Taipei
11529, Taiwan\\ and National Center  for   Theoretical   Sciences,
National Tsing   Hua   University,  Hsinchu  30013,   Taiwan}
\date{\today}

\begin{abstract}
We study the influence of linear friction on the vortex motion in a
non-viscous stratified compressible rotating media. Our method can
be applied to describe the complex behavior of a tropical cyclone
approaching  land. In particular, we show that several features of
the vortex in the atmosphere such as a significant track deflection,
sudden decay and intensification,
 can be
explained already  by means of the simplest two dimensional
barotropic model, which is a result of averaging over the height in
the primitive equations of air motion in the atmosphere.
 Our theoretical considerations are
in a good compliance with the experimental data. In contrast to
other models, where first the additional physically reasonable
simplifications are made, we deal with special solutions of the full
system. Our method is able to explain the phenomenon  of the cyclone
attracting to the land and interaction of the cyclone with an
island.
\end{abstract}

\numberwithin{equation}{section}
\def\sign{\mathop{\rm sgn}\nolimits}

\end{frontmatter}

There exists a lot of theoretical, numerical, and experimental
studies about the influence of land friction to the dynamics of
tropical cyclones (e.g.\cite{ChanWang} and references therein).
However, they contain sometimes contradictory results. For example,
some theoretical and numerical studies of the sensitivity of tropic
cyclones to the friction in axisymmetric models indicated that the
intensity decreases markedly as the drag coefficient increases
  \cite{Emanuel}, \cite{Craig}. At the same time,
 in \cite{Sensitivity1}, \cite{Sensitivity2},
 \cite{Luo},  the authors present a series of three-dimensional
convection-permitting numerical experiments in which the
intensification rate and intensity of the vortex increase with the
surface drag coefficient up to a certain threshold value and then
decrease. Further, the numerics made by MM5 (Fifth-Generation Penn
State/NCAR Mesoscale Model) have shown that  the phenomenon of
attraction of the cyclone to the land does exist \cite{Kuo}.
Moreover, many experimental works indicate that the circular air
motion in cyclone transforms into a convergent air flow during the
landfall \cite{Frederic}.

Of course,  the structure of a tropical cyclone is three-dimensional
and the processes of the moisture and heat transfer play an
important role in its formation. Nevertheless, we are going to show
that several very complicated three-dimensional phenomena can be
qualitatively explained  by a relatively simple mechanical
two-dimensional model where the topography is very crudely
parameterized by allowing the surface friction coefficient to vary.

We deal with a special class of solutions of the full system, which
is characterized by a linear profile of horizontal velocity. Under
this condition we show that a smooth  vortex in a material volume in
three-dimensional stratified compressible non-barotropic flow and
the simplest vortex in the two-dimensional barotropic model are
governed by the same nonlinear system of ODEs.

 We are going to show that the following
phenomena related to the cyclone motion can be explained by a rather
simple low parametric model analytically:  (i) a drastic
intensification of the cyclone and  a deflection of its track during
landfall; (ii) complex behavior of the cyclone during interacting
with the land (attraction
 and rounding); (iii) formation of converging flow starting from
 circular motion  inside the cyclone.


\section{Full model of gas dynamics adapted for atmosphere}\label{Sec1}
We consider the system of non-isentropic polytropic gas dynamics
equations in a uniformly rotating reference frame for unknown
functions $\rho\ge 0$, $p\ge 0$, $U=(U_1, U_2, U_3)$, $S$ (density,
pressure, velocity and entropy), in the presence of a horizontal dry
friction, namely
\begin{equation}\label{e1}
 \rho(\partial_t U+ (U,\nabla) U + l e_3\times U +\mu U_H + g e_3)=
-\nabla p,
\end{equation}
\begin{equation}\label{e2}
 \partial_t \rho + {\rm div}(\rho U)=0,
 \end{equation}
 \begin{equation}\label{e3}
 \partial_t S +(U,\nabla S)=0.
\end{equation}
   The functions depend on time $t$
and on point $x\in {\mathbb R}^3,$ $e_3=(0,0,1)$ is the "upward"
unit vector, $l$ is the Coriolis parameter, $\mu$ is the friction
coefficient, $g$ is the acceleration due to gravity (points in
$-e_3$ direction), $U_H=(U_1, U_2, 0)$, $\mu\ge 0$ is the friction
coefficient. The state equation is
\begin{equation}\label{e_state}
p=\rho^\gamma e^S,
\end{equation} where $\gamma>1$ is the adiabatic
exponent. This system is traditional, see e.g. \cite{Landau},
\cite{Pedloski}.

For our convenience taking into account \eqref{e_state} we write the
equation \eqref{e3} in terms of pressure:
\begin{equation}\label{e4}
 \partial_t p + (U, \nabla p) +\gamma \,p\,{\rm div} U=0. \end{equation}

Let us consider classical solutions of \eqref{e1}, \eqref{e2},
\eqref{e4}. For $\mu=0$, the system implies the conservation of mass
${\mathcal M}=\int\limits_{\Omega (t)}\rho \,dx$, momentum
${\mathcal P}=\int\limits_{\Omega (t)}\rho U\,dx$ and energy
$${\mathcal E}={\mathcal E}_k(t)+{\mathcal E}_p(t)=
\int\limits_{\Omega (t)}\left(\frac{\rho
|U|^2}{2}+\frac{1}{\gamma-1}p\right)\,dx,$$ inside a material volume
$\Omega (t)$, if we assume the hydrostatic balance
\begin{equation}\label{hydrostatic}
\partial_{x_3} p=-g \rho.
\end{equation}
To prove these conservation laws we apply the formula for the
derivative with respect to time of integral taken over a material
volume \cite{Ovs}, namely,
\begin{equation}\label{mov_vol}
\frac{d}{dt}\int\limits_{\Omega(t)}f(t,x) dx=
\int\limits_{\Omega(t)}\bigl(\partial_t f(t,x)+ {\rm div}
(f(t,x)\,U)\bigr)\,dx.\end{equation}

Let us introduce the following
 functionals:
 \begin{equation*}\label{G_F}
 G(t,x_3)=\frac{1}{2}\int \limits_{\Omega_H(t)}\rho|{
X_1}|^2\,d{ x_1}d{ x_2},\quad F_i(t,x_3)=\int\limits_{\Omega_H(t)}({
U, X}_i)\rho\,d{ x_1}d{ x_2},\end{equation*}
\begin{equation*}\label{Gxy1}
G_{x_1}(t,x_3)=\frac{1}{2}\int\limits_{\Omega_H(t)} \rho x_1^2 d{
x_1}d{ x_2},\quad
G_{x_2}(t,x_3)=\frac{1}{2}\int\limits_{\Omega_H(t)} \rho x_2^2 d{
x_1}d{ x_2},\end{equation*}
\begin{equation*}\label{Gxy2} G_{x_1
x_2}(t,x_3)=\frac{1}{2}\int\limits_{\Omega_H(t)} \rho x_1 x_2 \,d{
x_1}d{ x_2},
\end{equation*}
  where ${ X}_1=(x_1,x_2),\,{ X}_2=(x_2,-x_1),$ $\, i=1,2$, $\Omega_H(t)$
  is a section of $\Omega(t)$ at a fixed $x_3$. We note that  $G(t,x_3)>0$ and
$\Delta(t,x_3)=G_{x_1}G_{x_2}-G_{x_1 x_2}^2>0$ for nontrivial
solutions to \eqref{e1}, \eqref{e3}, \eqref{e4}.

Let us consider $l=\rm const\ge 0 $ and $\mu=\mu(x_3)$. Further, we
assume $ U_3=\Psi(t,x_3)$. The last assumption, \eqref{e1}, and
\eqref{hydrostatic} imply that $U_3(t,x_3)$ satisfies the Hopf
equation
$$\partial_t U_3+U_3 \partial_{x_3} U_3=0, \quad x_3>0.$$
It is well known that if the initial data $U_3(0,x_3)$ are not
increasing, then the solution will lose the smoothness within a
finite time. However, the  natural  conditions
$U_3(t,0)=U_3(t,+\infty)=0$ imply that the initial datum
$U_3(0,x_3)$ is not increasing everywhere. Therefore the only
possibility to consider a smooth solution for all $t>0$ is to set
$U_3=0$. This is the asymptotic of the solution containing shock
waves  as $t\to\infty$ on $(0,+\infty)$ for any initial data
$U_3(0,x_3)$ with the property $U_3(0,0)=U_3(0,+\infty)=0.$ Thus,
the assumption $U_3=0$ satisfies this limit case.

\begin{lemma}\label{lem1}
For the classical solutions of \eqref{e1}, \eqref{e2}, \eqref{e4}
with $U_3=0$ the following relations hold:
\begin{equation*}
\partial_t G=F_1,\quad \quad \partial_t F_2 =l F_1-\mu F_2,
\end{equation*}
\begin{equation*}
\partial_t F_1=2(\gamma-1){ E}_p+2 {E}_k - l F_2-\mu F_1,\end{equation*}
\begin{equation*}\label{2.1.7}
\partial_t {E}= -2\mu {E}_k,
\end{equation*}
where $E_k(t,x_3)=\frac{1}{2} \int\limits_{\Omega_H (t)}{\rho
|U_H|^2}\,d{ x_1}d{ x_2}\,$, $\,E_p(t,x_3)=
\frac{1}{\gamma-1}\,\int\limits_{\Omega_H (t)}p\,d{ x_1}d{ x_2}, $
$E=E_k+E_p$.
\end{lemma}

{\it Proof.} To prove the identities it is enough to apply formula
\eqref{mov_vol} with respect to the variables $x_2$ and $x_2$. For
example, taking into account \eqref{e2}, we get
$$\partial_t G=\frac{1}{2}\int\limits_{\Omega_H(t)}\partial_t\rho | X_1|^2\,dx_1
dx_2=$$$$ \int\limits_{\Omega_H(t)}\Bigl(-\frac{1}{2}\,{\rm
div}(\rho\, U)\, |{ X}_1|^2+\frac{1}{2}\,{\rm div}\bigl(\rho\, U_H\,
|X_1|^2)\bigl)\Bigr) \,\,dx_1 dx_2=$$$$=\int\limits_{\Omega_H(t)}
(X_1,U)\,\rho\,dx_1 dx_2=F_1.$$ The proof of other identities are
analogous. $\square$

Let us assume a special structure of velocity inside $\Omega(t)$.
Namely, we set \begin{equation}\label{linprof} U_H=Q_H {X_1},\quad
Q_H=\left(
\begin{array}{cc} a_H(t,x_3) & b_H(t,x_3)\\ c_H(t,x_3) &
d_H(t,x_3)\end{array}\right),\quad U_3=0.
\end{equation}
It is well known from experimental data that the profile of velocity
near the center of atmospherical vortex like tropical cyclone is
approximately linear \cite{Sheets}. Thus, we can consider the core
of the cyclone as $\Omega(t)$.

\begin{lemma}\label{lem2}
 For the
velocity \eqref{linprof} we have
$$\partial_t G_{x_1}=2a_H G_{x_1}+2b_H G_{x_1 x_2},\quad \partial_t G_{x_2}=2d_H G_{x_2}+2c_H G_{x_1
x_2},$$
$$ \partial_t G_{x_1 x_2}'=(a_H+d_H) G_{x_1 x_2}+ b_H G_{x_2}+ c_H
G_{x_1},
$$
$$
\partial_tE_p=-(\gamma-1)(a_H+d_H)E_p, \quad
\partial_t\Delta=2(a_H+d_H)\Delta.
$$
The potential energy $E_p$ is connected with $\Delta$ as
$$E_p(t,x_3)=E_p(0,x_3)\Delta^{(\gamma-1)/2}(0,x_3)\Delta^{(-\gamma+1)/2}(t,x_3).$$
\end{lemma} {\it Proof.} The proof is a direct computation with
taking into account formula \eqref{mov_vol}. The variable $x_3$
plays a role of parameter. $\square$

Let us introduce new functions
$$G_1={G_{x_1}}{\Delta^{-(\gamma+1)/2}},\quad
G_2={G_{x_2}}{\Delta^{-(\gamma+1)/2}},\quad G_3={G_{x_1
x_2}}{\Delta^{-(\gamma+1)/2}}.$$ Lemmas \ref{lem1} and \ref{lem2}
imply that  for the elements of the matrix $Q$ and $G_1,\,G_2,\,G_3$
can be obtained the following closed system of equations:
\begin{eqnarray}\label{sys_full}
\partial_t  G_1&=&((1-\gamma)a_H-(1+\gamma)d_H)G_1+2b_H G_3, \nonumber\\
\partial_t G_2&=&((1-\gamma)d_H-(1+\gamma)a_H)G_2+2c_HG_3,\nonumber\\
\partial_t G_3&=&c_HG_1+b_HG_2-\gamma(a_H+d_H)G_3,\nonumber\\
\partial_t  a_H &=&-a_H^2-b_H c_H+lc_H-\mu a_H-\mathcal K G_2,\\
\partial_t  b_H &=&-b_H(a_H+d_H)+l d_H-\mu b_H+\mathcal K G_3,\nonumber\\
\partial_t  c_H &=&-c_H(a_H+d_H)-la_H-\mu c_H+\mathcal K G_3,\nonumber\\
\partial_t  d_H &=&-d_H^2-b_H c_H-lb_H-\mu d_H-\mathcal K G_1,\nonumber
\end{eqnarray}
 with
$\mathcal K=\frac{\gamma-1}{2} E_p\Delta^{(\gamma-1)/2}|_{t=0}.$

Thus, \eqref{sys_full} describes the behavior of full 3D system of
 dynamics of stratified atmosphere near the center of atmospherical
 vortex.
 For every fixed level $x_3=\bar x_3$ we have its own motion of air
 and the dynamics of the whole material volume is determined by
  $a(0, x_3)$, $b(0, x_3),$ $ c(0, x_3)$, $d(0, x_3)$,
 $G_1(0, x_3)$, $G_2(0, x_3)$, $G_3(0, x_3)$.

Below we are going to show that the same system of ODE describes a
behavior of the simplest possible solution of barotropic
two-dimensional model of atmosphere.

\section{Bidimensional models of the atmosphere
}\label{Sec2}

 Since the horizontal scale in the atmospherical motion is much  larger than the
 vertical scale,  there exist many  approaches to simplifying  the
 model  \cite{Durran_Arakawa}, \cite{White}. Moreover, there is
a possibility of averaging over the height to hide vertical
processes and reduce the primitive system of equations
 to two space coordinates (see
\cite{Obukhov} for barotropic case and  \cite{Alishaev} for general
case).  Let us  recall shortly the procedure of averaging. Let
$\rho, \, U=(U_1,U_2,U_3),\, p, \,$ denote in the {\it
three-dimensional} density, velocity and pressure. Namely, all these
functions depend on $(t,x_1,x_2, x_3),$ $x_3\in\mathbb R_+$. Let us
introduce $\hat\phi$ and $\bar f$ to represent  an average of $\phi$
and $f$ over the height, respectively. The averaged values are
introduced as follows: $\displaystyle\hat\phi :=\int_0^\infty
\phi\,dx_3,\quad \bar f :=\frac{1}{\hat\rho}\int_0^\infty \rho
f\,dx_3$, where $\phi$ and $f$ are arbitrary functions, and denote
$\varrho(t,{ x_1, x_2})=\hat\rho, \,P(t,{  x_1, x_2})=\hat p,\,{\bf
U}(t,{ x_1, x_2})=(\bar U_1, \bar U_2).$ Moreover, the usual
adiabatic exponent, $\gamma$, is related to the ``two-dimensional"
adiabatic exponent $\gamma_H$ as follows:
$\gamma_H=\displaystyle\frac{2\gamma-1}{\gamma}<\gamma.$

 The impenetrability conditions are included in the model. We require the vanishing of
   derivatives of the velocity  on the
Earth surface and a sufficiently rapid decay for all thermodynamic
quantities as the vertical coordinate $x_3$ approaches to infinity.
In other words, the impenetrability conditions assure the
boundedness of the mass, energy, and momentum in the air column.
They also provide the necessary conditions for the convergence of
integrals.

If $l$ and $\mu$ are constant, the resulting two-dimensional system
consists of three equations for density $\varrho(t,{\bf x}),$
velocity ${\bf U}(t,{\bf x})$ and pressure $P(t, {\bf x})$, ${\bf
x}\in {\mathbb R}^2$:

\begin{align}\label{2d_U}
\begin{split}
 \varrho(\partial_t {\bf U} +  ({\bf U}\cdot \nabla ){\bf U} +
 {\mathcal L}{\bf U}) + \nabla P & = 0,
 \end{split}
\end{align}
\begin{align}
\label{2d_rho}
\begin{split}
\partial_t \varrho +  {\rm div} ( \varrho {\bf U}) &=0,
 \end{split}
\end{align}
\begin{align}
\label{2d_p}
\begin{split}
\partial_t P + ({\bf U}\cdot\nabla P)+\gamma_H P\,{\rm div}{\bf
U}&=0.
 \end{split}
\end{align}
Here $\mathcal L = l L + \mu I$,
$\quad L = \left(\begin{array}{cr} 0 & -1 \\
1 & 0
\end{array}\right)$,
$I$ is the identity matrix, $\gamma_H\in (1,2)$ .

We used this model with $\mu=0$ in our previous papers
\cite{RYH2010}, \cite{RYH2012}.

We can apply to \eqref{2d_U}-\eqref{2d_p} all considerations of
Sec.\ref{Sec1}.

However, for us it will be convenient to restrict ourselves to the
barotropic case, where $P={\mathcal C} \rho^\gamma,\, {\mathcal
C}={\rm const}$. Thus, the system under consideration can be reduced
to two equations \eqref{2d_U}, \eqref{2d_rho}.

We introduce a new variable $\pi=P^{\frac{\gamma-1}{\gamma}}$ and
get the system
\begin{equation}\label{2d_u}\begin{aligned}
\partial_t {\bf U} + ({\bf U} \cdot \nabla ){\bf U} +(l \, L \,+\mu \,I) {\bf u}+ c_0\, \nabla
\pi = 0,\\
\partial_t \pi +  (\nabla \pi \cdot {\bf U}) + (\gamma-1)\,\pi\,{\rm div}\,{\bf
U}\, =\,0, \end{aligned}
\end{equation}
with $c_0= \frac{\gamma}{\gamma-1} {\mathcal C}^{\frac{1}{\gamma}}.$

\subsection{ A class of exact solutions}\label{Exact_sol}

We consider a simple class of exact solutions which correspond to
the first terms of expansion of  $\pi$ at a critical point. Namely,
we  look for the solution in the form
\begin{equation}\label{u-form}{\bf U}(t,{\bf x})={ Q} {\bf x},
\qquad { Q}= \left(\begin{array}{cr} a(t) & b(t) \\
c(t)& d(t)\end{array}\right),\end{equation}
\begin{equation*}\pi(t,{\bf
x})=A(t)x_1^2+B(t)x_1x_2+C(t)x_2^2,
\end{equation*}
to get a closed ODE system for the components of the matrices $Q$
and $R=\left(\begin{array}{cr} A(t) & \frac12 B(t) \\
\frac12 B(t)&C(t),\end{array}\right) $:
\begin{equation}\begin{aligned}\label{m1}
\dot
R+ RQ + Q^T R+(\gamma-1){\bf tr}Q R =0,\\
\dot Q+Q^2+lLQ+\mu Q+2c_0 R=0.
 \end{aligned}
\end{equation}

The system of matrix equations consists of 7 nonlinear ODEs and has
a very complicated behavior. First of all, we discuss the behavior
of the system in the frictionless case.
 In fact, an analogous class of solutions was considered in
\cite{Ball} in another context.

It can be readily checked that system  \eqref{m1},  coincides
formally with \eqref{sys_full} at a fixed $x_3$, where ${ Q}=Q_H$,
$2 c_0=\mathcal K$, $A=G_2$, $B=-2G_3$, $C=G_1$. Nevertheless, the
nature of these systems is different and the components of solution
of \eqref{m1}  are not a result of integration of corresponding
components of solution of \eqref{sys_full} with respect to $x_3$. In
particular, the constants $\gamma$ and $\gamma_H$ are different for
these systems.

\subsection{A friction-free vortex ($\mu=0$), stability issue}

\subsubsection{Axisymmetric case \cite{RYH2010}}\label{s1} It is
easy to see that \eqref{m1} has a closed submanifold of solutions
having additional properties $a=d$, $c=-b$, $A=C$, $B=0$. These
solutions corresponds to the axisymmetric motion. Note that it is
the most interesting case related to the vortex in atmosphere.  Here
we get a system of 3 ODEs:
\begin{equation}\begin{aligned}\label{As0}\dot A+2\gamma
aA=0,\\ \dot a+a^2-b^2+lb+2c_0 A=0. \\
\dot b+2ab-la=0,
\end{aligned}
\end{equation} The functions
$a,\,b,\,A>0$ correspond to one half of divergence, one half of
vorticity and the fall of pressure in the center of vortex
respectively. The only nontrivial equilibrium point that relates to
a vortex motion is $$a=0,\, b=-c=b^*,\,
A=A*=\frac{b^*(b^*-l)}{2c_0}.$$ The center of vortex corresponds to
a domain of low pressure only if $A*>0$ (the motion is cyclonic).
This implies $b^*<0$ or $b^*>l$. Further, there exists one first
integral
\begin{equation}\label{mu0_integral}b=\dfrac
{l}{2}+ k A^{\frac{1}{\gamma}},\end{equation} where $k$ is a
constant \cite{RYH2010}. Thus, (\ref{As0}) can be reduced to the
following system:
\begin{equation*}
\label{A2s} \dot A=-2\gamma a A, \qquad \dot
a=-a^{2}-\dfrac{l^{2}}{4}+k^{2} A^{\frac{2}{\gamma}}-2 c_{0} A.
\end{equation*}
On the phase plane $\{(A,a),\,A>0\}$, there always exists a unique
 equilibrium  $(A^{*},a^{*})=(A_{0},0)$, stable in the Lyapunov sense
(a center), where $A_{0}$ is a positive root of equation
$$\dfrac{l^{2}}{4}+2c_{0}A=k^{2}A^{\frac{2}{\gamma}}.$$

\subsubsection{General case} As follows from \cite{Inviscid_damping}, the axisymmetric form of
2D vortex is stable with respect to asymmetric perturbations for the
solution to the incompressible Euler equations. Indeed, the
incompressibility condition implies $a(t)+d(t)=0$ and this reduces
the full system (\ref{m1}) to (\ref{As0}). As we have shown in
Sec.\ref{s1}, the equilibrium in this case  is stable for any $b^*$
and $l$.

Nevertheless, in the compressible case  this property does not hold
for arbitrary values of parameters.

\begin{theorem}\cite{RYTH}\label{TStable}
If
$$b^*<
\frac{1-\sqrt{2}}{2}\,l \quad \mbox{or} \quad
b^*>\frac{1+\sqrt{2}}{2}\,l>l,$$ then the equilibrium of system
(\ref{m1}) is unstable.
\end{theorem}

{\em Proof}.
 As one can readily check, the point $$a=d=0,\, b=-c=b^*,\,
A=C=A*=\frac{b^*(b^*-l)}{2c_0},\, B=0$$ is the only equilibrium of
the full system (\ref{m1}). It is the same point of equilibrium as
in the axisymmetric case (\ref{As0}). Nevertheless, in the symmetric
case this equilibrium is always stable in the Lyapunov sense,
whereas in the general case the situation is different. Indeed, the
eigenvalues of matrix corresponding to the linearization at the
equilibrium point are the following:
$$ \lambda_1=0,\quad \lambda_{2,3}=\pm
\sqrt{-(2(2-\gamma)b^*(b^*-l)+l^2)},$$  $$
\lambda_{4,5,6,7}=\pm\sqrt{
2}\,\sqrt{-l\left(b^*+\frac{l}{4}\right)\pm
\sqrt{\left(b^*+\frac{l}{2}\right)^2\left(\frac{l^2}{4}+b^*
l-(b^*)^2\right)}}.$$ Since $(2-\gamma)b^*(b^*-l)+l^2>0$ for
$\gamma\in(1,2)$, then $\Re(\lambda_{2,3})=0$. Eigenvalues
$\lambda_i,\,i=4,5,6,7$ have zero real part if and only if $b^*$
satisfies the following inequalities simultaneously: $
l(b^*+\frac{l}{4})\ge 0,\quad \frac{l^2}{4}+b^* l-(b^*)^2>0,$ $
l^2\left(b^*+\frac{l}{4}\right)^2>\left(b^*+\frac{l}{2}\right)^2\left(\frac{l^2}{4}+b^*
l-(b^*)^2\right),$ that is $b^*\in
\big[\frac{1-\sqrt{2}}{2}\,l,\,\frac{1+\sqrt{2}}{2}\,l \big]$. For
others values of $b^*$ the eigenvalues $\lambda_{4,5,6,7}=\pm \alpha
\pm i \beta, \,\alpha\ne 0, \beta \ne 0$, therefore there exist an
eigenvalue with a positive real part. Thus, the Lyapunov theorem
implies instability of the equilibrium for $b^*<
\frac{1-\sqrt{2}}{2}\,l$ and $b^*>\frac{1+\sqrt{2}}{2}\,l>l.$
$\square$

\begin{remark}
We notice that the full system (\ref{m1}) has the first integral
$$b-c-l= {\rm const}\,{\mathcal D}^{\frac{1}{2\gamma}},\quad
{\mathcal D}=4AC-B^2.$$ This reduces (\ref{m1}) to the system of 6
equations. If $b^*\in \Sigma$, $\Sigma=
\left(\frac{1-\sqrt{2}}{2}l,0\right)\cup
\left(l,\frac{1+\sqrt{2}}{2}l\right)$, then  the matrix,
corresponding to the system, linearized at the equilibrium, has 3
pairs of pure imaginary complex conjugate roots $\lambda_i,\,
i=2,\dots,7$ (for the range of parameters under consideration the
roots are simple). A study of stability in this case is extremely
complicated and we will not dwell here (see \cite{RYTH}).
\end{remark}

\subsection{Influence of the friction on an axisymmetric vortex}

The system of equations describing a vortex with a rotational
symmetry is the following:
\begin{equation}\begin{aligned}\label{Asm}\dot A+2\gamma
aA=0,\\
\dot a+a^2-b^2+lb+2c_0 A=-\mu a,\\
\dot b+2ab-la=-\mu b,\end{aligned}
\end{equation}
The solution to the equation has a complicated behavior,
nevertheless it is possible to study it analytically to a certain
extent.


\begin{theorem}\label{stability}
System (\ref{Asm}) has two equilibriums
$(a_{1}^{*},b_{1}^{*},A_{1}^{*})=(0,0,0)$ and
$(a_{2}^{*},b_{2}^{*},A_{2}^{*})=(-\mu,l,0)$, both are unstable.
\end{theorem}

{\em Proof}.
Indeed, 
the matrix of the system linearized at the point $(A_0,a_0,b_0)$ is
$$Q(A_0,a_0,b_0)=\begin{pmatrix}
 -2\gamma a_0&-2\gamma A_0 & 0  \\ -2c_{0}&
-2a_0-\mu & 2b_0-l  \\ 0& -2b_0+l & -2a_0-\mu
\end{pmatrix}.$$
The eigenvalues of $Q(0,-\mu,l)$ solve the equation
$$R(k)=(2\gamma\mu-k)((\mu-k)^{2}+l^2)=0.$$
The polynomial has a positive root.   This means instability of
equilibrium $(-\mu,l,0).$ The eigenvalues of $Q(0,0,0)$ are $(0,-\mu
\pm il)$, therefore the linearized theory does not give an answer to
the question about the stability or instability of zero equilibrium.
However in the critical case we can use the theory of \cite{Malkin},
Sec.4. Namely, we consider expansions into series $a(A)=a_1 A +
O(A^2)$ and $b(A)=b_1 A + O(A^2)$ as $A\to 0$. Then we substitute
the expansions into (\ref{Asm}) and get $a_1= -\frac{2\mu
c_0}{\mu^2+l^2}, $  $b_1= -\frac{2l c_0}{\mu^2+l^2}.$ Therefore
$\dot A = \frac{4 \gamma \mu c_0}{\mu^2+l^2}\,A^2+ O(A^2)$. This
implies  instability of zero equilibrium. $\square$
\begin{remark}
Theorem \ref{stability} implies that the the zero equilibrium of the
full system (\ref{m1}) is also unstable.
\end{remark}
\begin{theorem}\label{boundedness}
 The solutions to system
(\ref{Asm}) has no finite time blow up points at $t>0$ and the
following inequalities hold:
\begin{equation}\label{E_A}
\Lambda\le \frac{c_0}{\gamma-1}A + k_0 A^{\frac{1}{\gamma}} e^{-2\mu
t},
\end{equation}
\begin{equation}
\label{A_growth_estimates} A\le K_0 e^{\frac{2\mu\gamma
}{2-\gamma}t},
\end{equation}
where $\Lambda=\frac{a^2+b^2}{2}$, $\delta_0$, $K_0$ and $k_0$ are
positive constants depending only on initial data.
\end{theorem}
{\em Proof}. First of all we note that the first equation and two
latter equations of (\ref{Asm}) imply
\begin{equation}\label{Agams}
\dot A^{-1/\gamma}-2aA^{-1/\gamma}=0
\end{equation}
and
\begin{equation}\label{Es}
\dot \Lambda+2a\Lambda+2c_0 a A+2\mu \Lambda=0,
\end{equation}
respectively. Equations (\ref{Es}) and (\ref{Agams}) result
\begin{equation}\label{E_A_eq}
\frac{d}{dt}\,\left(\Lambda
A^{-\frac{1}{\gamma}}-\frac{c_0}{\gamma-1}A^{\frac{\gamma-1}{\gamma}}\right)=
-2\mu \Lambda A^{-\frac{1}{\gamma}}\le 0.
\end{equation}
From (\ref{E_A_eq}) we obtain (\ref{E_A}).

Let us prove  inequality \eqref{A_growth_estimates}. First of all we
note that inequality (\ref{E_A}) implies that there exists a
constant $\bar A,$ depending on initial data such that for $A>\bar
A$ we have
\begin{equation}\label{E_A_short}
\Lambda\le k_1 A
\end{equation}
with a positive constant $k_1$. Let us introduce a new variable
$W=(b-\frac{l}{2}) A^{-\frac{1}{\gamma}} e^{\mu t}.$ It is easy to
check that
\begin{equation}\label{Cs}
\frac{dW}{dt}=-\frac{l\mu}{2} A^{-\frac{1}{\gamma}} e^{\mu t} \le 0.
\end{equation}
As follows from (\ref{Cs}), $W(t)\le W(0).$ The second equation of
(\ref{Asm}) takes the form
\begin{equation}\label{aC}  a'=-a^{2}-\mu a+  W^2
A^{2/\gamma}e^{-2\gamma \mu t}-2 c_{0} A -\frac{l^2}{4}.
\end{equation}
First, we consider the cases  $l \mu=0,$ where $W(t)=W(0)$, and
$b(0)<\frac{l}{2}$ (or $W(0)<0$), for $l \mu\ne 0$. Then (\ref{aC})
and (\ref{E_A_short}) imply that for sufficiently large $A$ we have
\begin{equation}\label{a'_lower}
a'\ge W^2(0) A^{2/\gamma}e^{-2\gamma \mu t}-k_2 A, \quad k_2>0.
\end{equation}
Thus, if there exits an interval of $t$ such that the inequality
\begin{equation}\label{A_a'>0}
 A^{\frac{2-\gamma}{\gamma}}> k_3 e^{2\mu t}, \quad
 k_3=k_2/W^2(0),
\end{equation}
then for these $t$ the function $a(t)$ increases and, as follows
from the first equation of (\ref{Asm}), $A(t)$ decreases. Thus
$A(t)$  increases if and only if  inequality
(\ref{A_growth_estimates}), opposite to (\ref{A_a'>0}), holds.

 The last case is $l\mu \ne 0$, $b(0)\ge\frac{l}{2}$ or $W(0)\ge 0$.
 Due to \eqref{Cs} there are two possibilities: $W(t_*)<0$ for some
 $t_*>0$ or $W(t)>W_0={\rm const}\ge 0$ for all $t>0$. The first case
 can be reduced to the case $W(0)<0$ if we take $t_*$ as the initial
 moment of time. We note that $W(t)$ cannot be identically zero, this
 contradicts to the last equation of \eqref{Asm}.
If we assume $W_0>0$, the second possibility implies inequality
\eqref{a'_lower} with $W_0^2$ instead of $W^2(0)$. Thus, estimate
\eqref{A_growth_estimates} follows from the same reasoning. Let us
show that  $W_0$ does not vanish and our assumption is correct.
Indeed, from \eqref{E_A_short} we have
\begin{equation}\label{W}
W\le k_4 A^{\frac{\gamma-2}{2\gamma}} e^{\mu t},\quad k_4={\rm
const}>0,\end{equation}
 for sufficiently large $A$. Further, from
\eqref{Cs} and \eqref{W} we obtain
\begin{equation}\label{W1}
\frac{dW}{dt}\ge - k_5 W^{\frac{2}{2-\gamma}} e^{-\frac{\gamma \mu
t}{2-\gamma}},\quad k_5={\rm const}>0.\end{equation} We divide
variables in \eqref{W1} and  integrate.   After obvious estimates we
get
$$
W>\frac{W(0)}{(1+k_5
(W(0))^{\frac{\gamma}{2-\gamma}})^{\frac{2-\gamma}{\gamma}}}:=W_0>0.
$$
Thus, \eqref{A_growth_estimates} is proved. $\square$


\begin{remark}
For $\mu=0$ inequalities (\ref{A_growth_estimates}) and (\ref{E_A})
imply that the solution to system (\ref{As0}) is bounded for all
$t>0$ by a constant depending on the initial data.
\end{remark}
\begin{remark}
As follows from (\ref{Cs}), the value of  $ W(t)$ is constant for
$l\mu=0$. From the conservation of $W$ for $\mu=0$ we get the
integral (\ref{mu0_integral}).
\end{remark}
\begin{remark} \label{rem_bound} Although systems \eqref{m1} and \eqref{sys_full}
at a fixed $x_3$ are formally equivalent, we can obtain from
\eqref{sys_full} more information. Indeed, \eqref{m1} is considered
in the whole space ${\mathbb R}^2$ and it is not based on
conservation laws. In contrast, \eqref{sys_full} is considered in a
moving volume, where the balance of energy
 $E'(t)=-2\mu E_k(t)$ holds. Let us recall that the component $A(t)$
 of solution of \eqref{m1} corresponds to $G_1$ in the solution of
\eqref{sys_full}. Since we deal with the axisymmetric case,
$G_1=G_2=(G/2)^{-\gamma}$.   Let us note that if the velocity field
has the form \eqref{u-form}, $a=d,$ $c=-b$, then $E_k=(a^2+b^2)G\ge
0$, $E_p= \delta_1 G^{1-\gamma}\ge 0$, $E_k+E_p\le \delta_2$, where
$\delta_1$ and $\delta_2$ are positive constants, depending on
initial data. This implies $G\ge
\left({\delta_1}/{\delta_2}\right)^{\frac{1}{\gamma-1}}:=
\delta_3>0$. Thus, $G_1=(G/2)^{-\gamma}\le (\delta_3/2)^{-\gamma}$.
In terms of system \eqref{m1} this means that $A$ is bounded.
\end{remark}

\begin{figure}[h]
\begin{minipage}{0.4\columnwidth}
\centerline{\includegraphics[width=0.8\columnwidth]{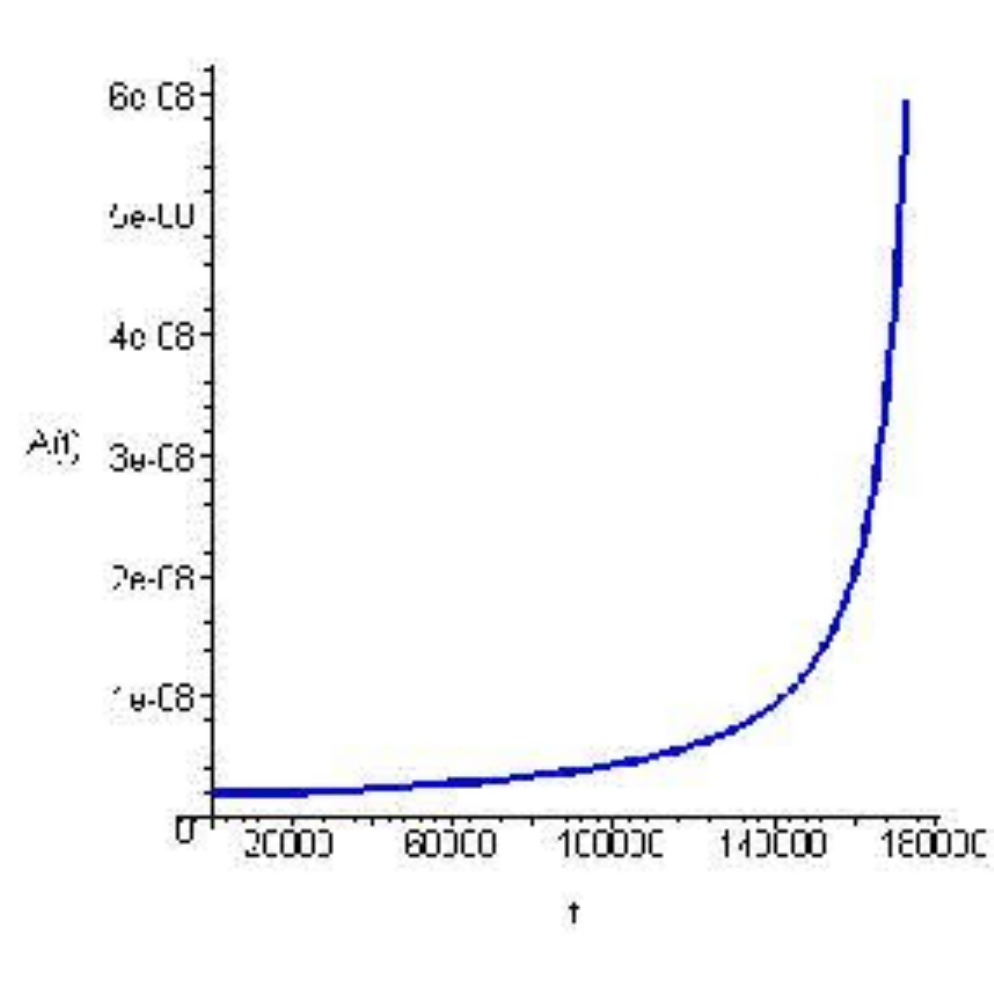}}
\end{minipage}
\begin{minipage}{0.4\columnwidth}
\centerline{\includegraphics[width=0.8\columnwidth]{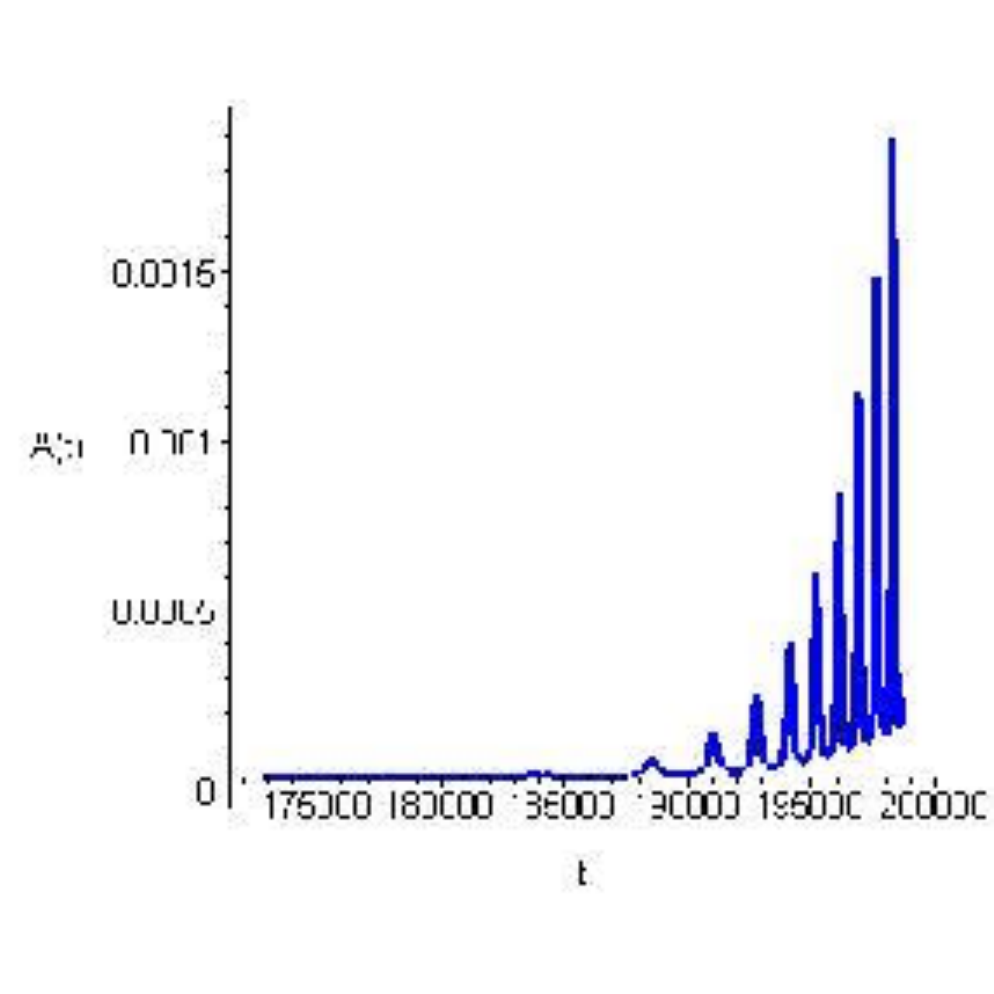}}
\end{minipage}
\caption{The intensification (left: 0 - 2 days) and destruction
(right: 2 - 2.3 days) of the vortex, $A(t)$ (intensity), the time is
in seconds.}\label{figure_AA}
\end{figure}

Figs.\ref{figure_AA}
shows the behavior of functions $A(t)$, related to the intensity for
different periods of time (0-2 days, 2-2.3 days). The parameters are
typical for geophysical vortex near its center. Namely, $c_0=0.1,\,
\gamma=\frac{9}{7},\, l=7.3\cdot 10^{-5}\, s^{-1}, \, b(0)=-5\cdot
10^{-6}\, s^{-1}, \, A(0) = 1.95\cdot 10^{-7} $ (solution to
equation $b^2(0)-l b(0)-2 c_0 A(0)=0$). The parameter $\mu$ is
$1\cdot 10^{-4} \, s^{-1}$. The initial data correspond to a steady
vortex for $\mu=0$. One can see that in the presence of the land
friction, the vortex first intensifies, after that a very fast
oscillations begin. These oscillations can be interpreted as a
destruction of the vortex. Moreover,  the increasing amplitude of
oscillations contradicts to  boundedness of $A$ in a moving volume
(Remark \ref{rem_bound}). Nevertheless, this contradicts only the
smoothness of solution in the moving volume and implies the
formation of frontal zone inside it.

\begin{figure}[h]
\begin{minipage}{0.45\columnwidth}
\centerline{\includegraphics[width=0.65\columnwidth]{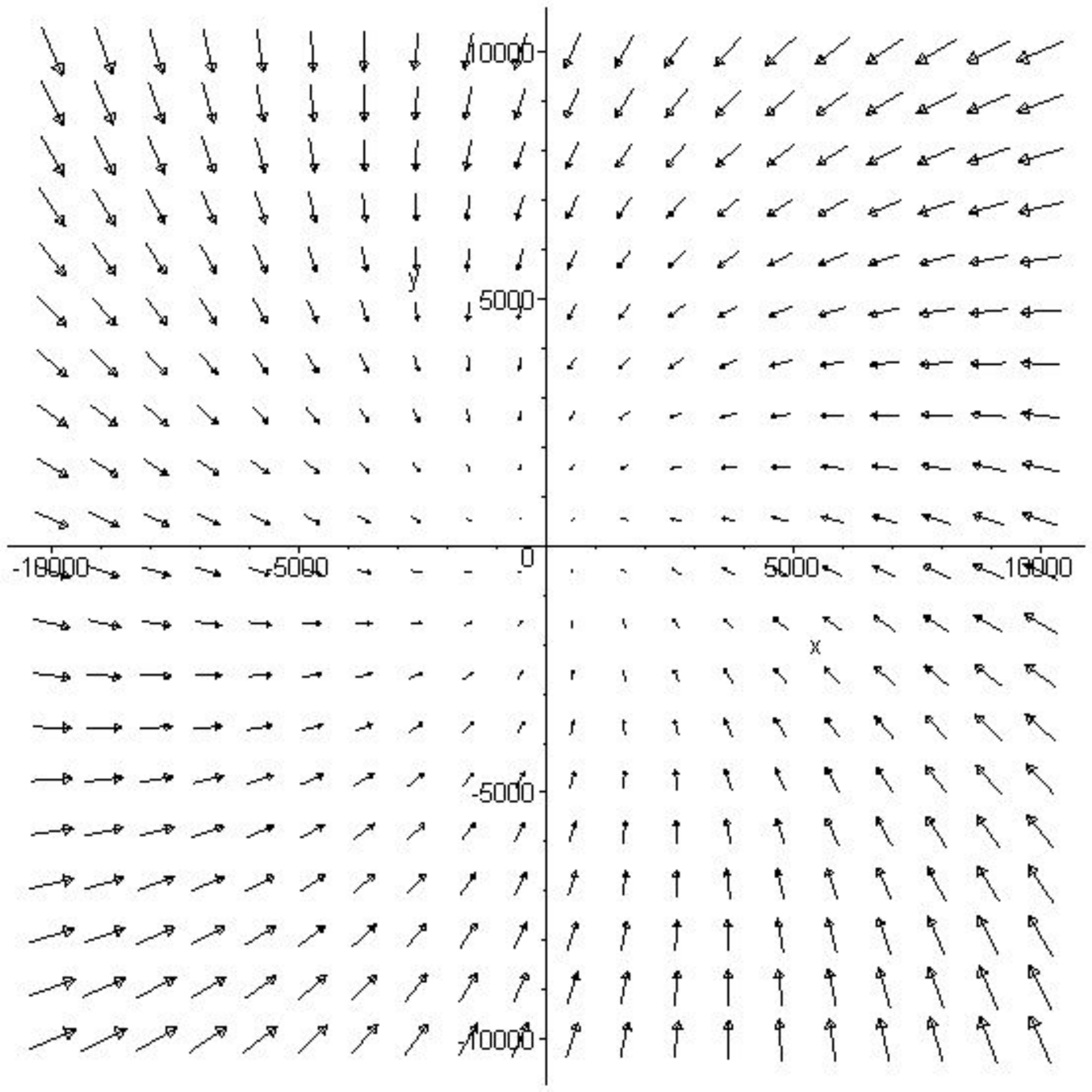}}
\end{minipage}
\begin{minipage}{0.45\columnwidth}
\centerline{\includegraphics[width=1.3\columnwidth]{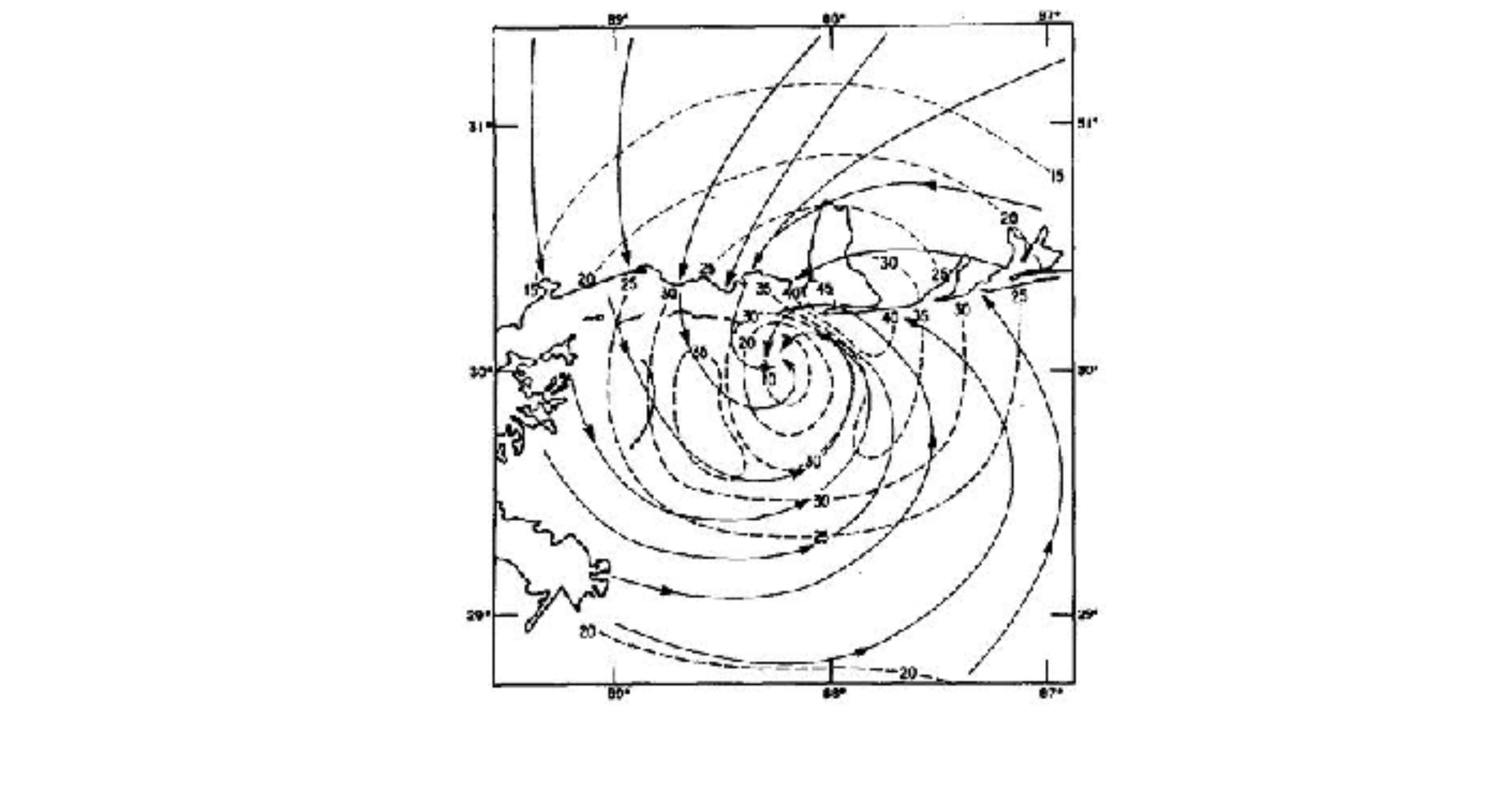}}
\end{minipage}
\caption{A convergent stream influenced by friction. Left: field of
wind (\ref{u-form}), axisymmetric case, built for $a(t)$ and $b(t)$
(see(\ref{As0})), the period of intensification. Right: streamline
analysis for the hurricane Frederic, experimental data
\cite{Frederic}}\label{figure_field}
\end{figure}

Further, Figs.\ref{figure_field} show the field of wind for a steady
vortex in the non-frictional case and the respective field
influenced by the constant surface friction for the period of
intensification of vortex, as in the left Fig.\ref{figure_AA}. We
can see a formation of a convergent stream. Fig.\ref{figure_field},
right, is taken from \cite{Frederic} and presents the experimental
evidence of the fact that the streamlines of the wind in a tropical
cyclone form the focal point near the landfall. Thus, in the frame
of the model it is possible to explain the formation of a convergent
stream from an axisymmetric steady state.

\begin{remark} The convergent inflow into
cyclone is a well known feature of rotating boundary layers in
general and can be explained be increasing of the Ekman pumping.
Nevertheless this phenomenon can be explained within a
two-dimensional non-viscous model.
\end{remark}



\section{Local and leading fields separation}\label{Sec3}

Let us  change the coordinate system of \eqref{2d_u} in such a way
that the origin of the new system ${\bf x}=(x_1,x_2)$ is located at
a point ${\bf X}(t)=(X_1(t),X_2(t))$. It is associated with the
center of vortex (here and below we use the lowercase letters for
$\bf x$ to denote the local coordinate system). Let ${\bf U}= {\bf
u}+{\bf V},$ where
 ${\bf V}(t)=(V_1(t),V_2(t))=(\dot X_1(t),\dot X_2(t))$.
Thus,
\begin{equation}\label{2d_sh_u}\begin{aligned}
\partial_t {\bf u} + ({\bf u} \cdot \nabla ){\bf u} +\\\nonumber
\dot {\bf V}+ (l \, L \,+\mu \,I) ({\bf u}+{\bf V})+ c_0\, \nabla
\pi = 0,\\
\partial_t \pi +  (\nabla \pi \cdot {\bf u}) + (\gamma-1)\,\pi\,{\rm div}\,{\bf
u}\, =\,0. \end{aligned}
\end{equation}

 Given a vector ${\bf V}$, the
trajectory of vortex can be found by integrating the system
\begin{equation}\label{X1X2}\dot X_1(t)=V_1(t),\qquad \dot
X_2(t)=V_2(t).\end{equation}

We assume that the pressure field can be separated into two parts
$$\pi=\pi_0(t,x_1,x_2)+\pi_1(t,x_1,x_2).$$
 The first field (we will call it local) is associated
with the vortex, the second field can be considered as a leading
one. In fact, the local field can be considered as a perturbation of
the leading field due to the vortex.   We impose a requirement
 $$\nabla \pi_0|_{{\bf
x}=0}=0.$$  Formally we can write
\begin{equation}\label{2d_sh_u}\begin{aligned} 
\big[\partial_t {\bf u} + ({\bf u} \cdot \nabla ){\bf u} + {\mathcal
L} {\bf u} + c_0\, \nabla \pi_0 \big]+\\ \big[\dot {\bf V}+
{\mathcal L} \, {\bf V}+ c_0\, \nabla \pi_1 \big]  =
0,\\\big[\partial_t \pi_0 + (\nabla \pi_0 \cdot {\bf u}) +
(\gamma-1)\,\pi_0\,{\rm div}\,{\bf u}\big]\,+\,\\\big[\partial_t
\pi_1 + (\nabla \pi_1 \cdot {\bf u}) + (\gamma-1)\,\pi_1\,{\rm
div}\,{\bf u}\,\big] =\,0,
\end{aligned}
\end{equation}
Let us denote
\begin{equation}\begin{aligned}\label{Q}
q =c_0\left[\nabla \pi_1(t,{\bf x})- \nabla\pi_1(t,{\bf
x})\Big|_{{\bf x}=0}\right].\end{aligned}\end{equation}

 If we solve separately the system for the local field
\begin{equation*}\begin{aligned} \label{u-nonsteady}
\partial_t {\bf u} + ({\bf u} \cdot \nabla ){\bf u} +  {\mathcal L}
\,{\bf u} + c_0\, \nabla \pi_0 + q =0,\\
\partial_t \pi_0 +  (\nabla \pi_0 \cdot {\bf u}) +
(\gamma-1)\,\pi_0\,{\rm div}\,{\bf u}=\,0,
\end{aligned}\end{equation*}
we get a  linear equation for $\pi_1$:
\begin{equation}\label{pi1}\begin{aligned}
\partial_t \pi_1 +  (\nabla \pi_1 \cdot {\bf u})+ \pi_1 {\rm div}{\bf u}= 0,
\end{aligned}
\end{equation}
which  can be solved for any initial condition $\pi_1(0,{\bf x})$.

Further, from  \eqref{2d_sh_u} we obtain
\begin{equation}\begin{aligned} \label{v}
\dot {\bf V}(t)+ {\mathcal L} {\bf V}(t) +c_0 \nabla \pi_1(t,{\bf
x})\Big|_{{\bf x}=0}= 0,\end{aligned}
\end{equation}
 then
\eqref{v} and \eqref{Q}  result
\begin{equation*}\begin{aligned} \label{xQ_exact}
\ddot {\bf X}(t)+ {\mathcal L} \dot{\bf X}(t) +c_0 \nabla
\pi_1(t,{\bf x})\Big|_{{\bf x}=0}= 0.\end{aligned}
\end{equation*}
If  $q=0$, then we obtain a complete separation of two processes and
the local field does not depend of the leading field. It is easy to
see that $q=0$ if and only if $\pi_1$ is linear with respect to the
space variables. If $q\ne 0$, however it is in some sense small, we
can talk about an approximate separation of processes; $|q|$ plays a
role of  measure of separability of the local and leading processes.

\section{Influence of friction  on the trajectory of vortex}

If the couple $({\bf u}, \pi_0)$ is found, then the position of the
vortex can be determined from  linear equations (\ref{pi1}) and
(\ref{Q}). Let $({\bf u}, \pi_0)$ be the exact solution considered
in Sec.\ref{Exact_sol}. As follows from (\ref{pi1}), if
$$\pi_1(0,x_1,x_2)= M(0) x_1 + N(0) x_2 +K(0), $$
then the initial field $\pi_1$ is linear with respect to the space
variables, that is
$$\pi_1(t,x_1,x_2)= M(t) x_1 + N(t) x_2 +K(t).$$ Thus, we
obtain the zero discrepancy term $q$ (i.e. the complete separation
of local and leading fields). The coefficients $M(t),$ $N(t)$ and
$K(t)$ can be found from the ODE system
\begin{equation}\label{MN}\begin{aligned}
        \dot {M} + (2 \gamma - 1) a M - b N = 0,\\
\dot{N} + (2 \gamma - 1) a N + b M = 0,\\
\dot K+2(\gamma-1){\bf tr} Q K=0.
\end{aligned}
\end{equation}
To obtain the trajectory of the center of vortex $(X_1(t), X_2(t))$,
we have to solve the system (\ref{v}), which can be reduced to
\begin{equation}\label{V1V2}\begin{aligned}
\dot{V}_1 - l V_2 + \mu V_1+c_0 M = 0,\\ \dot{V}_2 + l V_1 + \mu
V_1+ c_0 N = 0.
\end{aligned}
\end{equation}
Then the trajectory can be found from (\ref{X1X2}). The coefficients
$M(t)$ and $N(t)$ can be considered as a measure of intensity of the
leading field. It is natural to assume that they are so small that
the local field can be discerned in the leading field (see the
numerical examples from \cite{RYH2012}). The function $K(t)$ does
not influence  the trajectory.



\subsubsection{Constant coefficient of surface friction}
\begin{figure}[h]
\centerline{\includegraphics[width=0.4\columnwidth]{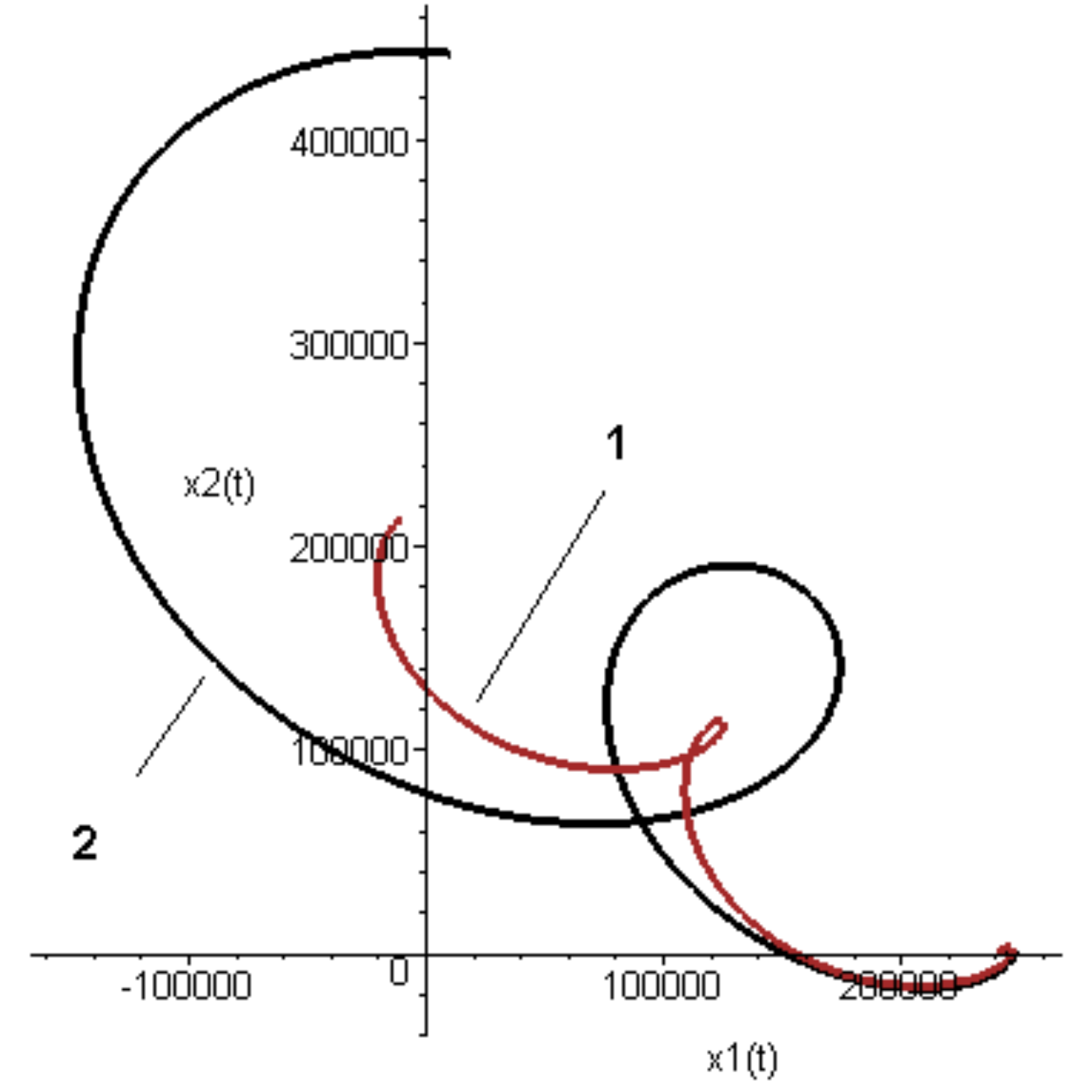}}
\caption{Influence of friction on the trajectories of "cyclone"
within 2 days: 1 - $\mu=0$, 2 - $\mu=2\cdot 10^{-5}$ m/s, initial
data are identical.}\label{land-see}
\end{figure}

 As we have shown,
the friction basically causes the intensification of vortex.
 At the same time, the trajectory of vortex changes according to the leading field
and initial velocity. It can shrink or amplify, formation of loops
is a typical behavior. Fig.\ref{land-see} shows  the position of
vortex, computed for the parameters corresponding to a tropical
cyclone within two days for $\mu=0$ and $\mu=2\cdot 10^{-5}{\rm
s}^{-1}$ respectively. The Coriolis parameter $l=7.3\times
10^{-5}\,{\rm s}^{-1}, $ that corresponds to the latitude
$30^\circ\,$ approximately, $c_0=0.1$ (appropriate dimension),
 $\gamma=\frac{9}{7}$ (recall that in the
procedure of averaging over the height, the value of heat ratio for
air changes). Initial data are the same for the both cases, namely,
$V_1(0)=V_2(0)=1$ m/s, $M(0)=N(0)=10^{-3}$ (appropriate dimension),
$b^*=-10^{-6} {\rm s}^{-1}$, initial condition corresponds to
equilibrium for $\mu=0.$

\subsubsection{ Influence of a land: interaction of vortex with the
"island"}


 The most intriguing phenomenon is the interaction of the cyclone
with the land. There are a lot of experimental and numerical
evidences showing that the cyclone "feels" the land; it can be
attracted to the shore, but sometimes it "avoids" the shore, on the
contrary \cite{Yuan}, \cite{TangWang1}, \cite{TangWang2}. Now we are
going to show that the approximation of trajectory made by the ODE
system (\ref{Asm}),(\ref{MN}),(\ref{V1V2}), (\ref{X1X2}) can
reproduce this complicated behavior. The coefficient of surface
friction $\mu$ in this experiments is a function of space variables
$X_1, X_2$. It changes from zero (sea surface) to some constant
value $\mu_0$ ("island"). Let  $\mu(X_1, X_2)=\omega
(X_1,X_2)\mu_0$,
\begin{equation*}\begin{aligned}
\omega (X_1,X_2)= \prod_{k=1}^2\left(\arctan\left(\frac{X_k(t)+\bar
x_k}{\sigma}\right)- \arctan\left(\frac{X_k(t)-\bar
x_k}{\sigma}\right)\right),
\end{aligned}\end{equation*} where
$\sigma$ is a very small constant. The square $[-\bar x_1,\bar
x_1]\times [-\bar x_2,\bar x_2]$ corresponds to the island, the
motion begins over the sea.

\begin{figure}[h!]
\begin{minipage}{0.45\columnwidth}
\centerline{\includegraphics[width=1.5\columnwidth]{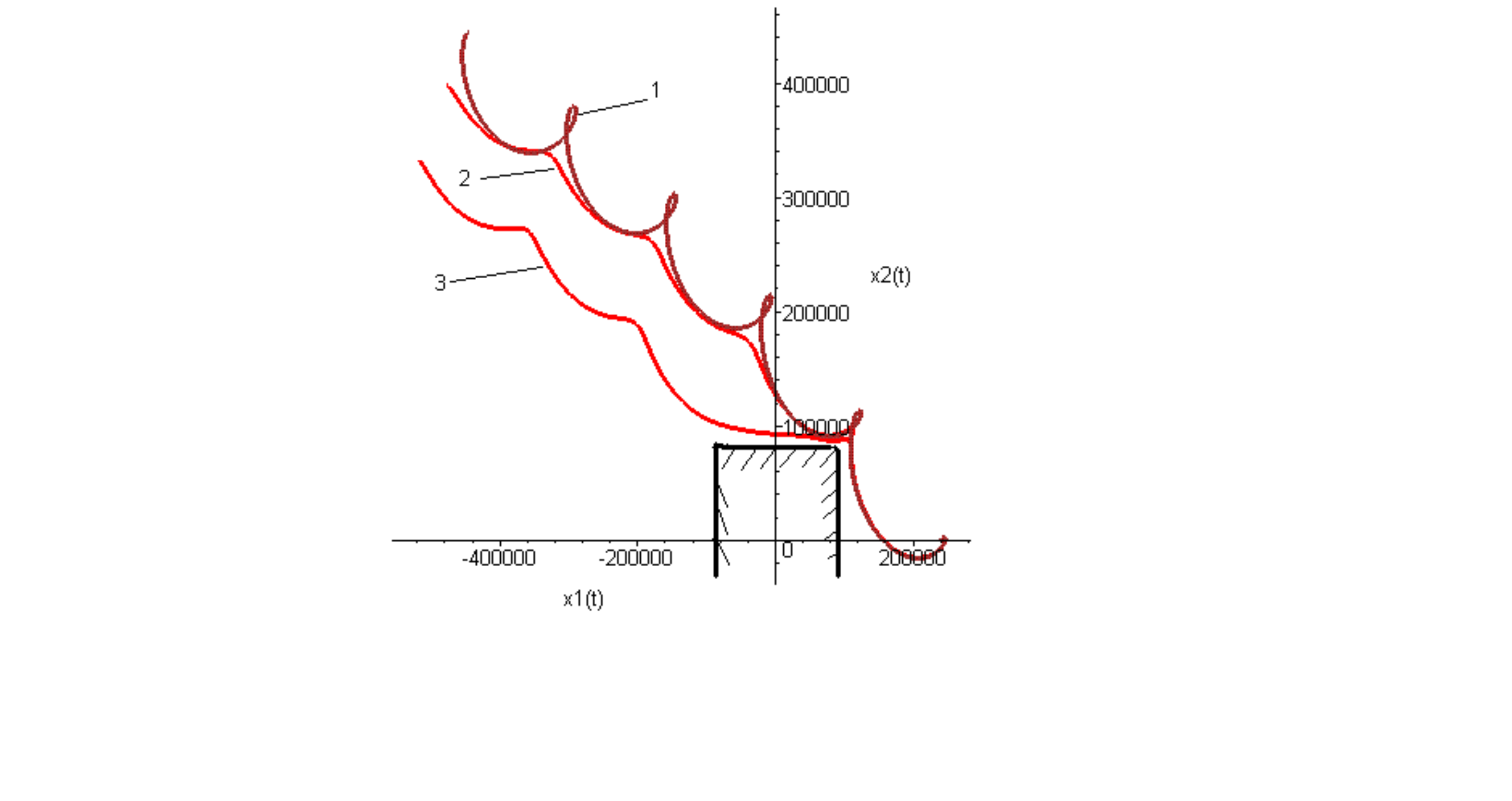}}
\end{minipage}
\begin{minipage}{0.45\columnwidth}
\centerline{\includegraphics[width=1.5\columnwidth]{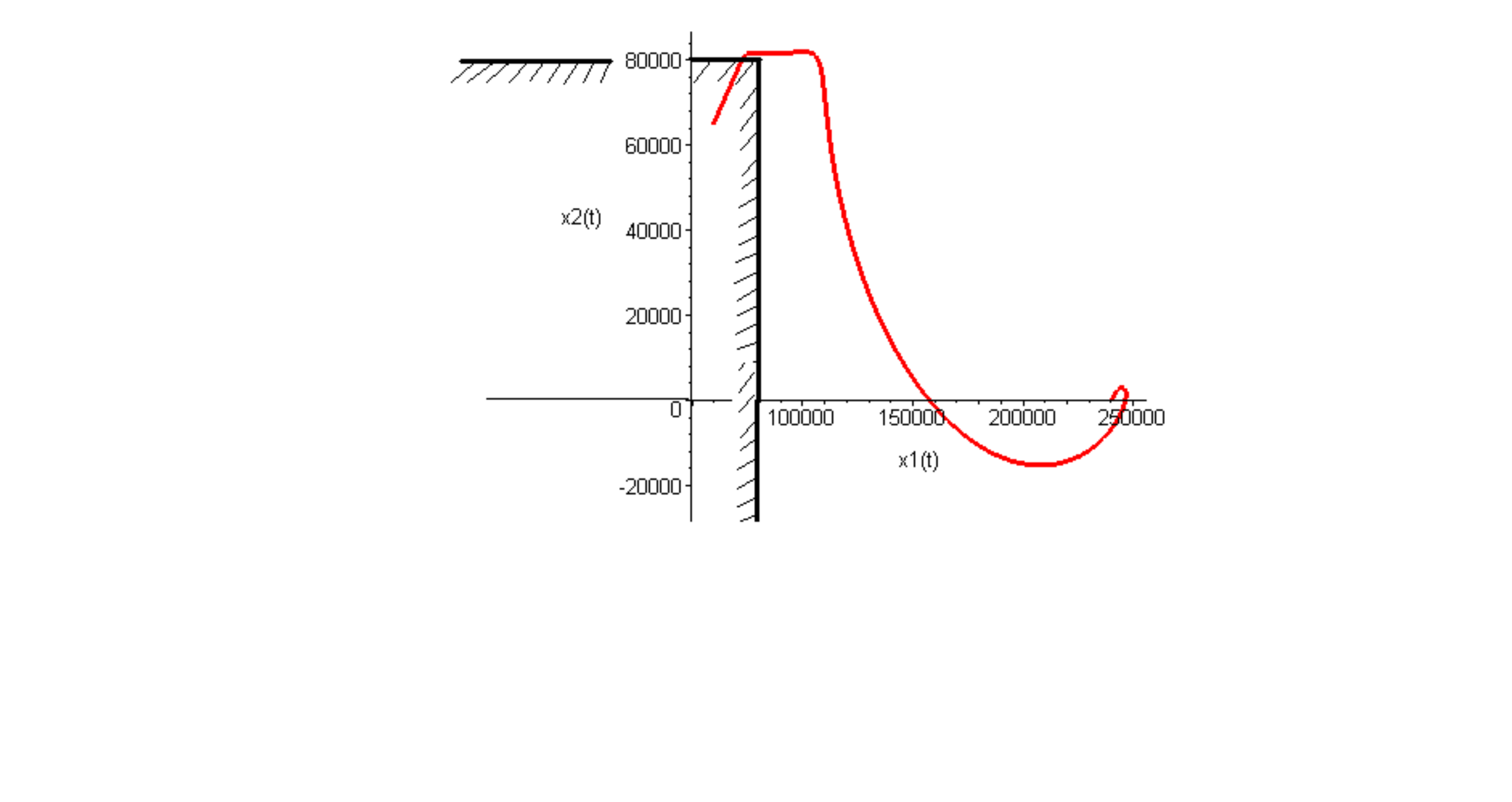}}
\end{minipage}
\caption{Attraction of cyclone to the island depends on the
coefficient of friction. Computations made for 5 days,
$V_1(0)=V_2(0)=1 \,{\rm m/s},\,$ $M(0)=N(0)=10^{-3}$ (appropriate
dimension) Left: 1 - $\mu_0=0$, 2 - $\mu_0=10^{-4}$, 3 - $\mu_0=2$
(${\rm s}^{-1}$).}\label{attraction_trajectory}
\end{figure}

\begin{figure}[h!]
\centerline{\includegraphics[width=0.3\columnwidth]{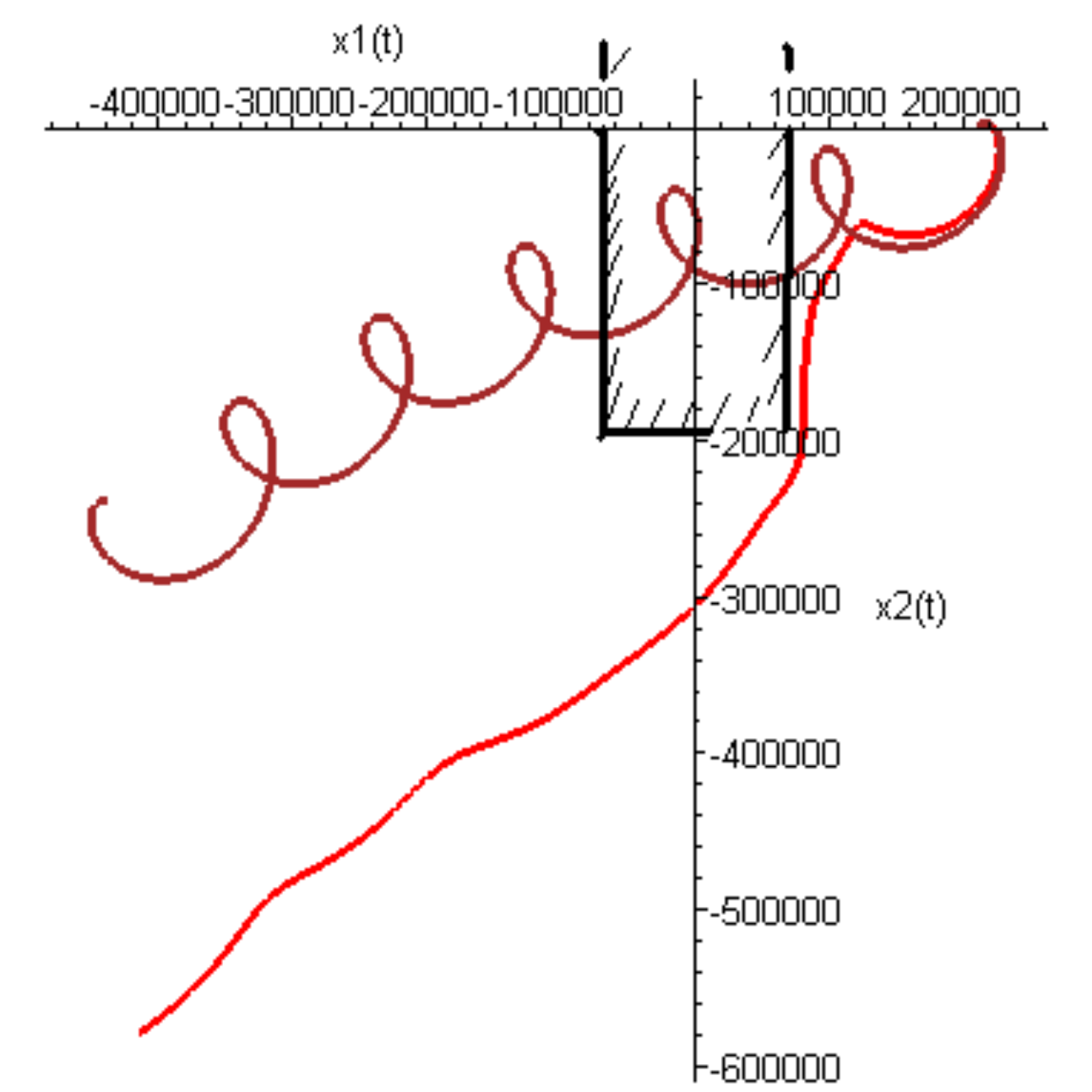}}
\caption{Avoiding of the shore: comparison with the motion over sea,
dimensions of the island are $[-70, 70]\times [-200, 200] \,(km)$.
$V_1(0)=V_2(0)=1 \,{\rm m/s},\,$ $M(0)=-10^{-4},\,N(0)=10^{-3}$
(appropriate dimension),
 computations made for 8 days. 1 - $\mu_0=0$, 2 - $\mu_0=10^{-3}$ (${\rm s}^{-1}$).}\label{avoiding}
\end{figure}

Fig.\ref{attraction_trajectory} shows that the attraction to the
lands increases with the roughness of the land.
Fig.\ref{attraction_parameter} shows the difference of divergence
and intensity for different roughness of the island.
Fig.\ref{avoiding} shows that for some initial conditions, the
cyclone can avoid the island.

\begin{figure}[h]\label{attraction_parameter}
\begin{minipage}{0.45\columnwidth}
\centerline{\includegraphics[width=0.8\columnwidth]{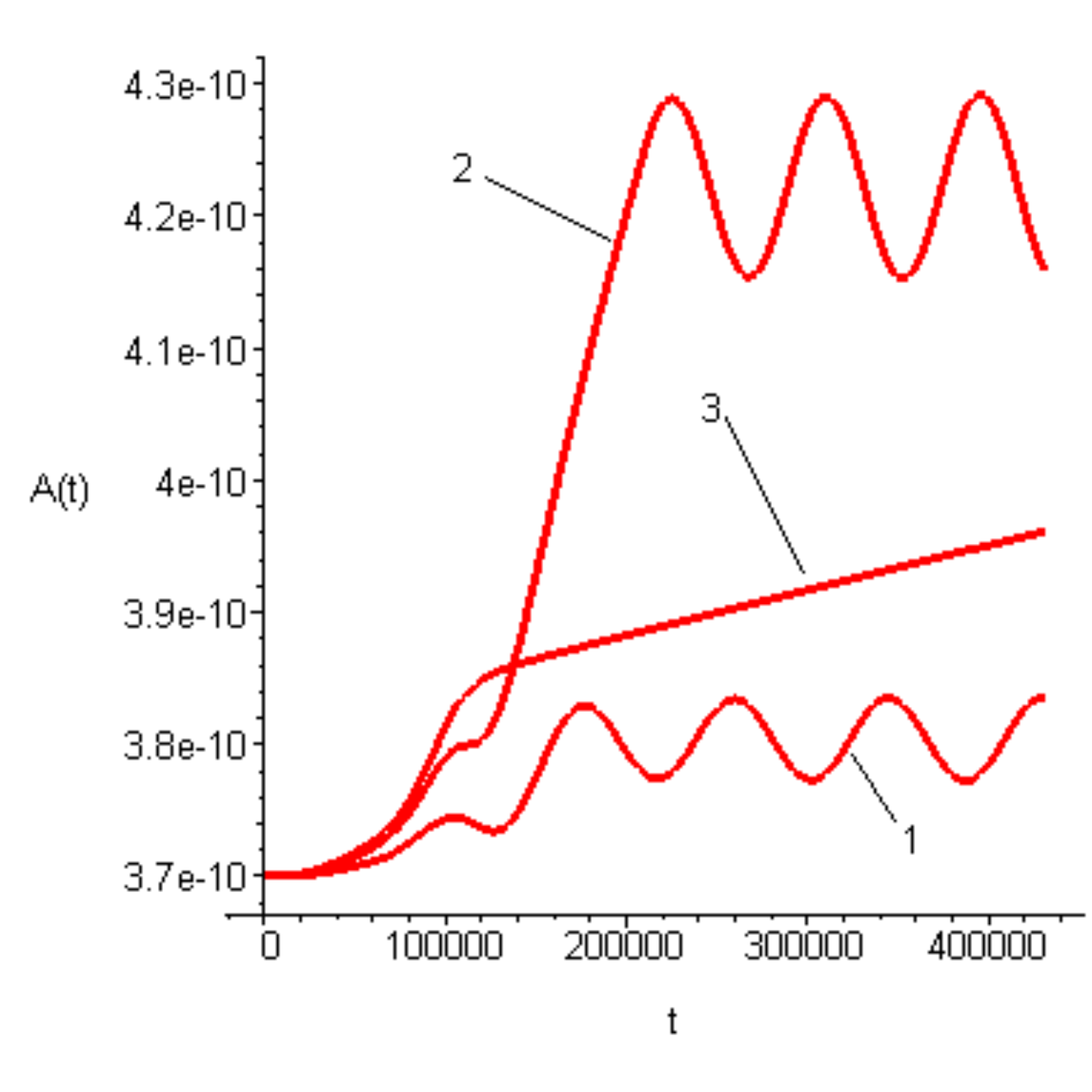}}
\end{minipage}
\begin{minipage}{0.45\columnwidth}
\centerline{\includegraphics[width=0.8\columnwidth]{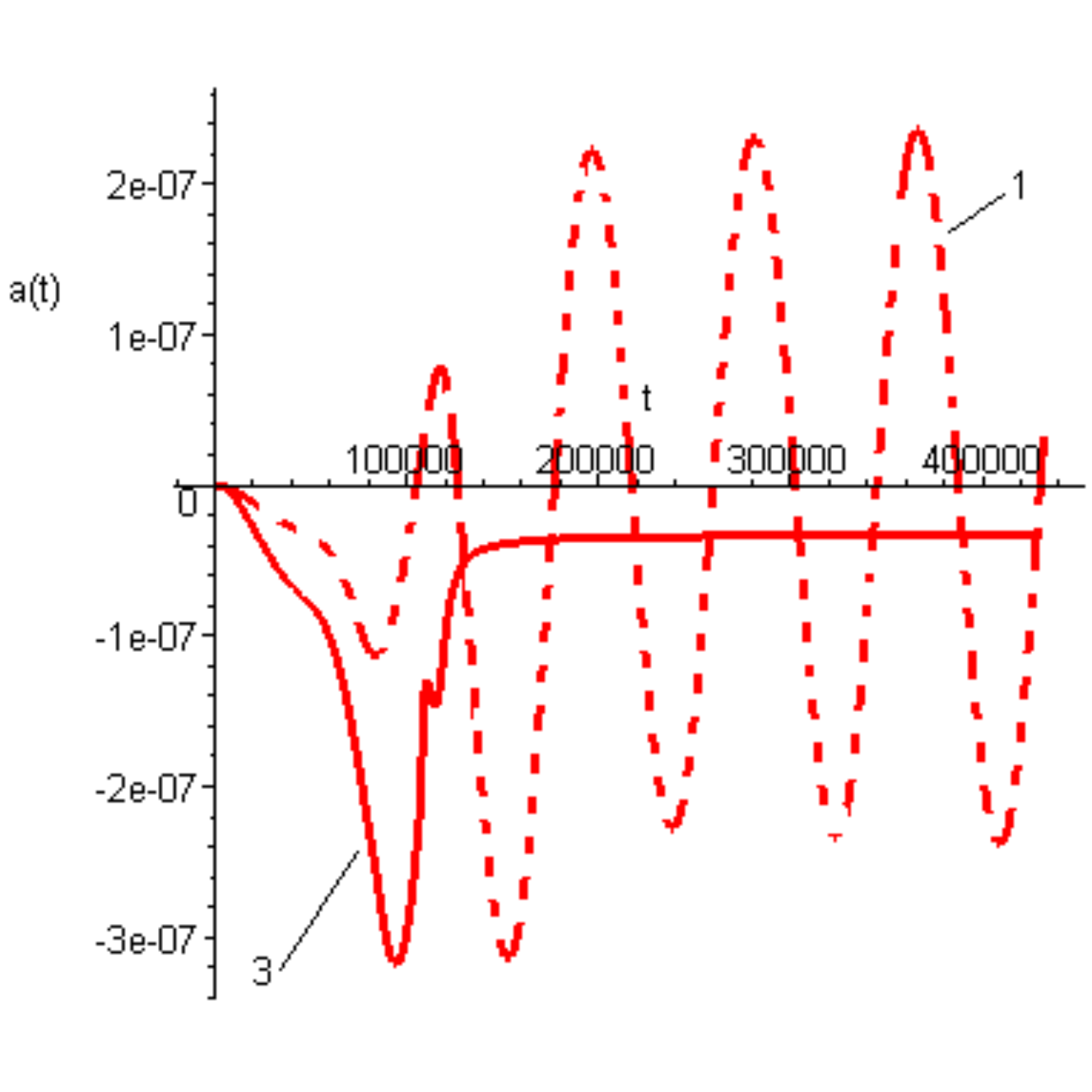}}
\end{minipage}
\caption{Intensity (left) and divergence (right) of the vortex. 1 -
$\mu_0=10^{-5}$, 2 - $\mu_0=2\cdot 10^{-5}$, 3 - $\mu_0=2.5 \cdot
10^{-5}$ (${\rm s}^{-1}$). }\label{attraction_parameter}
\end{figure}



\section{Conclusion}

To study different properties of vortices in a compressible medium
(e.g. atmosphere), we analyze a special class of smooth motions
characterized by a linear profile of the horizontal velocity. It is
well known that the velocity has this property near the center of
vortex (e.g.\cite{Sheets}). We consider two models: the primitive 3D
model
 of atmosphere and 2D model obtained from it  by  the
standard averaging procedure.

 In the latter model we make additional
assumption about barotropicity of the process to obtain a system
having an exact solution in the form of the first terms of the
Taylor expansion in the center of  vortex. Thus, we get a system of
two equations. We could obtain the same result by assuming that the
process is isochoric, i.e. the pressure is proportional to the
temperature. We study the influence of the linear  friction on the
behavior of vortex.


We show that in the case of atmosphere, where the vertical motion
satisfies the hydrostatic balance and the horizontal and vertical
motion are somewhat separated,
 the vortex behavior can be
described by the same system of ODEs as the vortex in the 2D
barotropic model. The vertical coordinate can be considered  as a
parameter.

First of all, we consider a steady axisymmeric vortex in the
friction-free case and show that in contrast to the incompressible
case, there exist some parameters such that the vortex is unstable
with respect to small perturbations of initial symmetry in the class
of solution with a linear profile of velocity.

 Further, we prove that the vortex always loses the stability when
 the surface friction coefficient is constant.
 We note that the flutter instability (a blowing-up vibrational
motion) can be induced by dry friction in mechanical systems which
would be stable without frictional forces (e.g.\cite{Bigoni}).

Finally, we study an axisymmetric vortex in the case of constant
surface friction both analytically and numerically
 and show that the
complicated features of the vortex (e.g. a sudden decay and further
intensification)  can be explained by our model. Moreover, we have
shown that the phenomenon of interaction of the tropical typhoon
with land can be quite realistically explained by a relatively
simple ODE system, where the effect of topography is modeled by
variable surface friction coefficient.

The aim of our study is to show that many interesting effect of the
atmospherical vortex motion can be qualitatively explained already
by a very simple mechanical model. Of course, we do not pretend to
claim that the trajectories of real tropical cyclones can be
entirely explained by this model. There are at least three reasons
for this. First of all, the real atmospheric vortex is localized and
has structure (\ref{2d_sh_u}) only in a vicinity of its center. The
exact solution with linear profile of velocity gives the exact
separation of local and leading fields only if the leading field is
also linear with respect to the space variables. In this case the
discrepancy $q= 0$ (see \eqref{Q}) and the position of the center of
vortex can be computed from the nonlinear system of ODEs
(\ref{Asm}),(\ref{MN}),(\ref{V1V2}), (\ref{X1X2}) exactly.  If the
velocity and pressure have a more realistic localized structure,
then the discrepancy $q\ne 0$. Nevertheless, as follows from
(\ref{pi1}), $q$ remains small until the divergence of the velocity
of the local field is small. As we have shown in \cite{RYH2012}, for
the case $\mu=0$, the difference between position of the center
obtained from the ODE system  and
 the position of
the localized vortex obtained from direct numerical computations can
be very small within several days. The second reason is that for big
velocities the drag friction coefficient $\mu$ depends on velocity
itself (thus, the  surface stress parametrization should be
quadratic). Nevertheless, as one can easily see from \eqref{e1}, the
quadratic drag friction together with the geostrophic condition
$\rho l L { U_H}+\nabla_H p=0$ ($\nabla_H$ stands for the horizontal
gradient) lead to a fast singularity formation for any realistic
profile of velocity. Therefore, this kind of friction worsens  the
smoothness of the solution significantly. The third reason also
relates to our assumption about the smoothness of solution. Indeed,
as we have shown in Remark \ref{rem_bound}, the fact of the envelope
of oscillations  growing with time contradicts to the conservation
of energy for the moving volume. This contradiction can be removed
if we assume a formation of shock wave within the volume. Therefore,
we can say that  the "theoretical" trajectory of vortex is close to
a real one only for some period of time depending on initial data.
The computations made for realistic parameters suggest that this
interval is less than one day. Therefore the rising oscillations
that one can see in the solution of the nonlinear system of ODE
cannot develop.

All these questions about correspondence of theoretical and real
vortex can be solved only numerically.
 The influence of different kinds of friction on the trajectory of a
localized vortex and  comparisons   with the trajectory computed
from (\ref{Asm}),(\ref{MN}),(\ref{V1V2}), (\ref{X1X2}) are the
issues of our future researches.


\begin{center}{ACKNOWLEDGMENTS}\end{center}

OSR thanks Dr.A.Akhmetzhanov and M.Turzynsky for a valuable
discussion.

JLY was supported by the grant of NSC 102-2115-M-126-003 and thanks
Academia Sinica for supporting the summer visit. OSR was supported
by Mathematics Research Promotion Center, MOST of Taiwan and RFBR
Project Nr.12-01-00308.


\end{document}